\documentstyle[epsfig,11pt]{article}
\title{\bf BFKL pomeron propagator in the external field of the nucleus}
\author{M.A.Braun, A.N.Tarasov  \\
S.Peterburg State University, Russia}
\begin{document}
\maketitle
\input epsf
\def\beq{\begin{equation}}
\def\eeq{\end{equation}}
\def\phid{\Phi^{\dagger}}
\def\pd{\partial}
\def\lra{\leftrightarrow}
\def\balpha{\bar{\alpha}}
\medskip

{\bf Abstract}

It is shown by numerical calculations that the convoluted QCD
pomeron propagator in the external field created by a solution of the
Balitsky-Kovchegov equation in the nuclear matter vanishes at high
rapidities. This may open a possibility to apply the perturbative approach
for the calculation of pomeron loops.

\section{Introduction}
In the QCD, in the limit of large number of colours,
strong interaction at high energies is mediated by the exchange of
BFKL pomerons, which interact via their splitting and fusion.
In the quasi-classical
approximation  for photon (hadron)-nucleus scattering the relevant
tree (fan) diagrams
are summed by the well-known Balitsky-Kovchegov (BK) evolution equation
 ~\cite{bal,kov,bra1}. For nucleus-nucleus scattering appropriate
 quasi-classical equations were derived in ~\cite{bra2,bra3}.
In both cases pomeron loops were neglected.
This  approximation can be
justified if the parameter $\gamma=\lambda\exp{\Delta_P y}$ is small,
with $y$ the rapidity and $\Delta_P$ and $\lambda$ the pomeron intercept
and triple pomeron coupling.
Then for a large nuclear target, such that $A^{1/3}\gamma\sim 1$,
the tree diagrams indeed give the dominant contribution and loops
can be dropped. However with the growth of $y$ the loop
contribution becomes not small and this approximation breaks down.

Direct calculation of the loop contribution seems to
be a formidable task  for the non-local BFKL pomeron.
Simplest loops have been studied
in several papers for purely hadronic scattering ~\cite{peschansky,
bartels, bra4}. In particular in ~\cite{bra4} it has been found
that pomeron loops become essential already at rapidities of the
order 10$\div 15$. They shift the position of the pomeron pole
to the complex plane and thus lead to oscillations in cross-sections.
However with the growth of energy loop contributions begin to dominate
and one needs to sum all of them. There have been many attempts to do this
in the framework of the so-called reaction-diffusion formulation of the
QCD dynamics and the following correspondence with the statistical approach
~\cite{iancu,AHM,levlub,lev,marquet,levmil} (see also a review
~\cite{soyez} and references therein).
Unfortunately concrete results could be  obtained only
with very crude approximations  for the basic BFKL interaction and the
stochastical noise in the statistical formulation.
The conclusions of different groups are incomplete and contradictory.
So in ~\cite{levmil} it was found that the geometric scaling following
from the BK equation was preserved with loops taken into account,
although going to the black disc limit was much slower.
On the contrary in papers based on the analogy with statistical physics
(see ~\cite{marquet,soyez}) it was argued that the BK scaling was changed
to the so called
diffusive scaling (with an extra $\sqrt{y}$ in the denominator of the
argument) but the speed of achieving the black disk limit was essentially
unchanged.

In our previous study of pomeron loops ~\cite{BT}
we considered a much simpler model with the local supercritical pomeron
in the Regge-Gribov formalism.
Instead of trying to solve
the model for the purely hadronic scattering we considered
the hadron-nucleus scattering and propagation of the pomeron inside
the heavy nucleus target. Moreover to avoid using numerical
solution of the tree diagrams contribution with
diffusion in the impact parameter, we concentrated on the case of
a constant nuclear density, which allowed to start with the known
analytical solutions. We have found that the
nuclear surrounding  transforms the pomeron from the supercritical
one with intercept $\epsilon>0$ to a subcritical one with the intercept
$-\epsilon$. Then Regge cuts, corresponding to loop diagrams, start at
branch points located to the left of the pomeron pole and their contribution
is subdominant at high energies. As a result, the theory acquired the
properties similar to the Regge-Gribov one with a subcritical pomeron
and allows for application of the perturbation theory. In~\cite{BT}
we expressed our hopes that a similar phenomenon might occur in the
QCD with BFKL pomerons.

In this note we demonstrate that such hopes are possibly founded.
We consider the pomeron propagator in the nuclear field
 and give arguments that, similar to the local
Regge-Gribov case,
it vanishes at large rapidity distances. We stress that at present
we are unable to give the full proof for this behaviour. Our study is
based on numerical calculations.
This makes us to choose  a relatively small
subset of initial conditions out of the complete set
necessary for the study of the pomeron propagator.
Our numerical results show that, with the chosen set of initial conditions,
this pomeron  propagator vanishes at large rapidity distances.
This result  may serve as a starting point  for the study of loops
and  can be directly applied for
double inclusive cross-sections in the nucleus-nucleus scattering.

\section{Local pomeron in the nuclear field}
To make clear our motivation and goals we start with reviewing our treatment
of a simpler model, the local reggeon field theory (LRFT) with a supercritical
pomeron and triple
pomeron interaction. The following  passage to the  non-local BFKL pomeron
will not introduce ideologically new moments, although greatly complicates
the technique.

The LRFT model is based on two pomeron fields
$\Phi(y,b)$ and $\phid(y,b)$ depending on the rapidity $y$ and
impact parameter $b$, with a Lagrangian density ~\cite{schwimmer}
 \beq
L=L_0+\lambda\phid\Phi(\Phi+\phid)+g\rho\phid.
\label{lagden}
\eeq
Here the free Lagrangian density is
\beq
L_0=\phid\Big(\frac{1}{2}\stackrel{\leftrightarrow}{\partial_y}
-\alpha'\nabla_b^2+\epsilon\Big)\Phi\equiv \phid S\phi,
\label{freel}
\eeq
where $\epsilon$ is the intercept minus unity and $\alpha'$ is the slope.
The  source term describing interaction with the nuclear target at low
energies is
\beq
g\rho(y,b)=gAT_A(b)\delta(y),
\label{lext}
\eeq
where $g$ is the pomeron-nucleon coupling constant and $T(b)$ the
profile function of the nucleus.
For a supercritical pomeron $\epsilon>0$ and $\lambda<0$.

Solution of the classical equation of motion for field $\phid$ is $\phid_0=0$
and the classical equation for $\Phi$ becomes
\beq
\partial_y\Phi=-\alpha'\nabla_b^2\Phi
+\epsilon\Phi+\lambda{\Phi}^2,
\label{eqphi}
\eeq
with an initial condition
\beq
\Phi(y=0)=gT_A(b).
\label{iniphi}
\eeq
Equation (\ref{eqphi}) describes evolution of the pomeron field in
rapidity and its diffusion in the impact parameter inside the nucleus.
We denote the solution of the classical equation of motion
(\ref{eqphi}) with the initial condition (\ref{iniphi}) as
$\Phi_0(y,b)$.

\begin{figure}
\hspace*{4 cm}
\epsfig{file=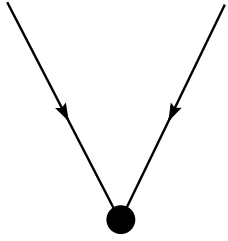,width=3 cm}
\caption{The new vertex for two-pomeron annihilation, which appears
after the shift in field $\Phi$}
\label{fig1}
\end{figure}

\begin{figure}
\hspace*{4 cm}
\epsfig{file=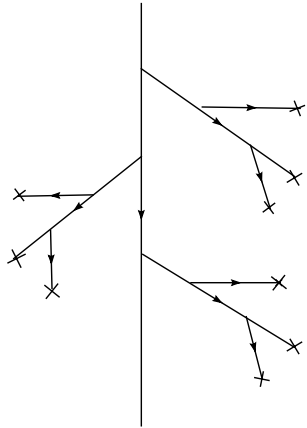,width=4 cm}
\caption{Diagrams summed into the pomeron propagator in the nuclear field}
\label{fig2}
\end{figure}

To go beyond the classical approximation and thus study loops one makes
a shift in the quantum field $\Phi$:
\beq
\Phi(y,b)=\Phi_0(y,b)+\Phi_1(y,b)
\label{shift}
\eeq
and reinterprets the theory in terms of quantum fields
$\Phi_1$ and $\phid$.
In the  Lagrangian terms linear in $\phid$
vanish due to the equation of motion for $\Phi$ and
we obtain
\beq
L=\Phi_1(S+2\lambda \Phi_0)\phid+\lambda\Phi_0{\phid}^2+
\lambda\Phi_1\phid (\Phi_1+\phid).
\eeq
This Lagrangian corresponds to a theory in the vacuum with the
pomeron propagator  in the external field $f(b,y)=2\lambda\Phi_0(y,b)$
\beq
P =-(S+2\lambda\Phi_0)^{-1},
\label{prop}
\eeq
the standard triple interaction
and an extra interaction described by the term $\lambda\Phi_0{\phid}^2$.
This new
interaction corresponds to  transition of a pair of pomerons into the vacuum
at point $(y,b)$ with a vertex $\lambda\Phi_0(y,b)$, see Fig. \ref{fig1}.
The propagator $P$ in the external field is described by a sum of diagrams
shown in Fig. \ref{fig2}.

\begin{figure}
\hspace*{5 cm}
\epsfig{file=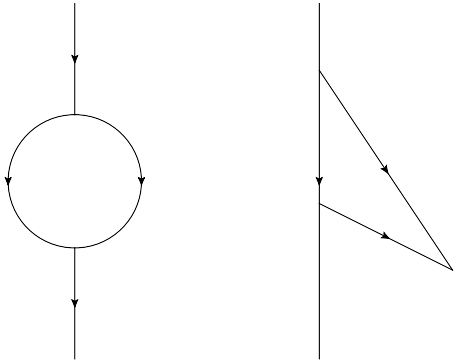,width=6 cm}
\caption{Simplest loop diagrams for the pomeron Green function}
\label{fig3}
\end{figure}

Loops  can be formed both by the standard interaction
and the new one. In the latter case they are to be accompanied by at
least a pair of standard interactions.
Simplest loops in the Green function
are illustrated in Fig. \ref{fig3}, where it is assumed that the propagators
are all taken in the nuclear field $f$.
Note that a loop formed by the standard
interaction has the lowest order $\lambda^2/\alpha'$
and requires renormalization in this model.
A loop formed by the new interaction has the lowest order
 $\lambda^3/\alpha'$ and
is finite.

\begin{figure}
\hspace*{5 cm}
\epsfig{file=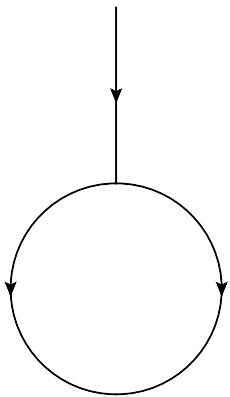,width=3 cm}
\caption{The lowest order diagram
for the quantum correction to the scattering amplitude}
\label{fig4}
\end{figure}

The quantum part of the amplitude is obtained as a tadpole
$g<\Phi_1(y,b)>$. The simplest
diagram for it contains one loop and is shown in Fig. \ref{fig4}.

Propagator  (\ref{prop}) in the external field
$f(y,b)=2\lambda\Phi_0(y,b)$  satisfies the equation
\beq
\frac{\partial P(y,b|y'b')}{\partial y}=
(\epsilon-\alpha'\nabla_b^2) P(y,b|y',b')+
f(y,b)P(y,b|y',b')
\label{eqprop}
\eeq
with the boundary conditions
\beq
P(y,b|y',b')=0,\ \ y-y'<0,\ \ P(y',b|y',b')=\delta^2(b-b').
\label{inigfb}
\eeq
In the general case  propagator $P$ can only be calculated
numerically, just as the external field $f(y,b)$. Its analytic form
can be found in two cases ~\cite{BT}: if the slope $\alpha'=0$ or
if one is studying the nuclear matter with the profile function $T_A(b)=T_0$
independent of $b$. In the second case, relevant for our following studies,
the nuclear field becomes independent of $b$ (but dependent on $y$)
and the propagator in this field takes the form
\beq
P(y,b|y',b')=\frac{1}{4\pi\alpha' (y-y')}\,\frac{p^2(y')}{p^2(y)}\,
e^{\epsilon (y-y')-\frac{(b-b')^2}{4\alpha'(y-y')}},
\label{prop1}
\eeq
where
\beq
p(y)= 1-\frac{\lambda g T_0}{\epsilon}\Big(e^{y\epsilon}-1\Big).
\eeq
Remarkably at $y\to\infty$ this propagator vanishes as $e^{-(y-y')\epsilon}$.
This means that contributions from the loops will be small at large $y$,
so that one can calculate them by the standard perturbation approach.

\section{Interacting BFKL pomerons}
Now we pass to our main subject, the theory of interacting BFKL pomerons.
By its structure it is quite similar to the considered LRFT model.
However now the pomeron, apart of the impact parameter $b$, depends on the
relative transversal distance $r$
between the two reggeized gluons out of which it
is formed. So it is described by two fields $\Phi(y,r_1,r_2)$ and
$\phid(y,r_1,r_2)$ depending on rapidity $y$ and transverse coordinates
of the reggeized gluons $r_1$ and $r_2$.
The  Lagrangian has a general structure similar to
(\ref{lagden}):
\beq
L=L_0+L_I+L_E.
\eeq
Here the free Lagrangian is
\beq
L_0=\int d^2r_1d^2r_2\phid
\nabla_1^2\nabla_2^2\Big(\frac{\pd}{\pd y}+H_{BFKL}\Big)\Phi,
\eeq
where $H_{BFKL}$ is the Hamiltonian for the BFKL pomeron ~\cite{lipatov}.
The triple pomeron interaction is given by
\beq
L_I=\frac{2\alpha_s^2N_c}{\pi}
\int \frac{d^2r_1d^2r_2d^2r_3}{r_{12}^2r_{23}^2r_{31}^2}
\phid(y,r_1,r_2)\phid(y, r_2,r_3)K_{31}\Phi(y,r_3,r_1)
+\Big(\Phi\lra\phid \Big),
\eeq
where $r_{ik}=r_i-r_k$ and
\beq
K_{31}=r_{31}^4\nabla_3^2\nabla_1^2.
\eeq
The interaction with the nuclear target is
\beq
L_E=-\int d^2r_1d^2r_2\phid J,
\eeq
where
\beq
J(y,r_1,r_2)=gAT_A(b)\rho(r_{12})\delta(y),
\eeq
$T_A$ is as before the profile function of the nucleus
and $\rho$ is the colour density of the nucleon.

Our idea is to repeat the procedure applied to the local pomeron as described
in the preceding section. First one solves the classical equation for the
fields. Its solution for $\Phi$ is a sum of all tree diagrams with the
pomeron splitting in two. This solution satisfies the BK
equation  and although not known analytically can  be found
numerically. At the next step we calculate the
quantum correction to this solution (pomeron loops). To this aim we shift the
field $\Phi$ with a non-zero classical value $\Phi_0$ by this value,
passing to the new quantum field $\Phi_1=\Phi-\Phi_0$, exactly as for the
LRFT, and study the theory in terms of fields $\Phi_1$ and $\phid$.
This leads to a change in the Lagrangian quite similar to what has been
described in the preceding section. In the new theory the propagator
is to be taken in the nuclear field. This is achieved by substituting
one of the two fields $\Phi$ by its classical value
$\Phi_0$ in the part of interaction quadratic in $\Phi$. Also in the new
theory there appears a new interaction term,
obtained when in the part of interaction linear in $\Phi$ one substitutes
$\Phi$ by its classical value.

In this paper we are not attempting at the calculation of loops.
We limit ourselves with the starting point of this program: studying
the pomeron propagator in the described nuclear field. Our aim is
to understand if this propagator vanishes at large rapidities
instead of growing in the
absence of the field, as it occurs in the LRFT. If it is indeed so then
one can hope to calculate loop contributions  in the nuclear
surrounding perturbatively. Otherwise this surrounding  is
useless for loop calculations.
The BFKL propagator $P(y,r_1,r_2;y',r'_1,r'_2)$ in the nuclear
field  satisfies the equation
\[
\frac{\pd P(y,r_1,r_2)}{\pd y}
=\frac{\balpha}{2\pi}\int d^2r_3\frac{r_{12}^2}{r_{13}^2r_{23}^2}\]\beq
\Big(P(y,r_1,r_3)+P(y,r_2,r_3)-
P(y,r_1,r_2)-\Phi(y,r_1,r_3)P(y,r_2,r_3)-\Phi(y,r_2,r_3)P(y,r_1,r_3)\Big),
\label{eqq}
\eeq
where $\bar{\alpha}=\alpha_sN_c/\pi$ and
we suppressed the second three variables
coordinates $y',r'_1,r'_2$ on which the
equation does not depend. The dependence on them follows from the initial
condition:
\beq
P(y=y',r_1,r_2;y',r'_1,r'_2)=
\nabla_1^{-2}\nabla_2^{-2}\delta^2(r_1-r'_1)\delta^2(r_2-r'_2).
\label{inicond}
\eeq
Function $\Phi(y,r_1,r_2)$ in this equation is the solution of the
BK equation
for the sum of all fan diagrams in the nucleus.
It satisfies a non-linear equation
\beq
\frac{\partial \Phi (y,r_1,r_2)}{\partial y}=
\frac{\bar{\alpha}}{2\pi}\int d^2r_3\frac{r_{12}^2}{r_{13}^2r_{23}^2}
\Big(\Phi(y,r_1,r_3)+\Phi (y,r_2,r_3)-
\Phi(y,r_1,r_2)-\Phi(y,r_1,r_3)\Phi(y,r_2,r_3)\Big)
\label{ebk}
\eeq
with the initial condition appropriate for a given nucleus
\beq
\Phi(y=0,r_1,r_2)\equiv \Phi_0(r_1,r_2)=1-e^{-T_A(b)\rho(r_{12})}.
\eeq
We remind that  $b=(r_1+r_2)/2$, $T_A$ is the conventional profile
function of the nucleus
and $\rho$ is the colour density of the nucleon.

We shall again study the simple case of the nuclear matter when
$T_A(b)=T_0$ is independent of $b$. Then from the structure of Eq. (\ref{ebk})
it follows that $\Phi$ is a function of only the relative distance between
the gluons $\Phi(y,r_1,r_2)=\Phi(y,r_{12})$.
Since the study is only possible numerically,
to avoid using the singular initial condition,
we shall consider a convolution of $P(y,r_1,r_2;y',r'_1,r'_2)$ with an arbitrary
function $\nabla_1^2\nabla_2^2\psi(r_1,r_2)$:
\beq
P(y,r_1,r_2)=\int d^2r_1'd^2r_2'P(y,r_1,r_2;y',r_1',r_2')\nabla_1^2\nabla_2^2
\psi(r_1',r_2').
\eeq
This convolution satisfies the same equation as the propagator itself
but at $y=y'$ we have
\beq
P(y=y',r_1,r_2)=\psi(r_1,r_2).
\label{ini}
\eeq
Obviously the properties of the propagator can be studied taking a full
set of functions $\psi(r_1,r_2)$.

Note that in the following, for simplicity, we mostly set $y'=0$.
The dependence on the choice of $y'$ will be studied for the forward case
and shown to preserve our main conclusions.


\section{Forward case}
\subsection{Main equations}
Calculation become especially simple for the forward case when not only
$\Phi$ but also the convoluted propagator $P$ depend only on the relative
distance between the gluons in the pomeron. In this case the equation
for $P$ simplifies to
\beq
\frac{\partial P(y,x)}{\partial y}=
\frac{\bar{\alpha}}{2\pi}\int d^2x_1\frac{x^2}{x_1^2x_2^2}
\Big(P(y,x_1)+P(y,x_2)-P(y,x)-2\Phi(y,x_1)P(y,x_2)\Big).
\label{eqf}
\eeq
Here we denoted $x=r_{12}$, $x_1=r_{13}$ and $x_2=r_{32}$ with $x_1+x_2=x$.
In the same notation the nuclear field $\Phi$ satisfies
\beq
\frac{\partial\Phi(y,x)}{\partial y}=
\frac{\bar{\alpha}}{2\pi}\int d^2x_1\frac{x^2}{x_1^2x_2^2}
\Big(\Phi(y,x_1)+\Phi(y,x_2)-\Phi(y,x)-\Phi(y,x_1)\Phi(y,x_2)\Big),
\label{ebkf}
\eeq

We have to solve this pair of equations with the
initial condition for $\Phi(y=0,x)$  fixed by the
properties of the nuclear medium, and  the initial condition (\ref{ini})
for $P(y,x)$ taken arbitrary, since we are interested in the
propagator in a given nuclear surrounding.

Equation (\ref{eqf}) is a linear equation for $P(y,x)$ in contrast to the
BK equation. At $y\to\infty$ $\Phi(y,x)\to 1$
independent of the chosen initial condition. One may think that at
$y\to\infty$ the behaviour of $P(y,x)$ can be derived from the asymptotic
equation
\[
\frac{\partial P(y,x)}{\partial y}\Big|_{y\to\infty}=
\frac{\bar{\alpha}}{2\pi}\int d^2x_1\frac{x^2}{x_1^2x_2^2}
\Big(P(y,x_1)+P(y,x_2)-P(y,x)-2P(y,x_2)\Big)\]\beq=
-P(y,x)\frac{\bar{\alpha}}{2\pi}\int d^2x_1\frac{x^2}{x_1^2x_2^2}.
\label{eq11}
\eeq
However the integral on the right-hand side has become divergent
(although it converges at finite $y$). This means that the limit
$y\to\infty$ is more delicate and cannot be taken under the sign of integral
over $x_1$. And indeed we shall see by numerical calculations that
the behaviour of the solution at $y\to\infty$ is not solely determined
by the limiting value of $\Phi(y,x)$ but depends on its behaviour
at finite $y$.

For numerical studies  both the BK equation and
linear equation (\ref{eqf}) in the momentum space are more convenient.
Introducing
\beq
\phi(y,x)=\frac{\Phi(y,x)}{x^2},\ \ p(y,x)=\frac{P(y,x)}{x^2}
\label{transmom}
\eeq
and then passing to the momentum space we obtain the following equations for
$\phi(y,k)$ and $p(y,k)$
\beq
\frac{\partial\phi(y,k)}{\partial y}=
-\bar{\alpha}\Big(H_{BFKL}\phi(y,k)+\phi^2(y,k)\Big)
\label{bkmom}
\eeq
and
\beq
\frac{\partial p(y,k)}{\partial y}=
-\bar{\alpha}\Big(H_{BFKL}+2\phi(y,k)\Big)p(y,k),
\label{eq1mom}
\eeq
where
\beq
H_{BFKL}=\ln k^2+\ln x^2-2(\psi(1)+\ln 2).
\eeq
To study the behaviour of the propagator in the external field $\phi$
one has to solve this pair of equations with the initial conditions
\beq
\phi(y,k)_{y=0}=\phi_0(k),\ \ p(y,k)_{y=0}=p_0(k),
\label{inimom}
\eeq
with some fixed $\phi_0$ and for a complete set of function $p_0(k)$.

\subsection{Numerical studies. Momentum space}
We have set up a program which simultaneously solves the
pair of equations (\ref{bkmom}) and (\ref{eq1mom}) for a given pair
of initial conditions (\ref{inimom}) by the Runge-Kutta method.
For the BK evolution we
have fixed the initial condition as
\beq
       \phi_0(k)=-(1/2){\rm Ei}(-k^2/0.3657)
\label{inphi}
\eeq
used in our previous calculations. We have limited our momenta
in the interval $10^{-8}<k<10^8$ and divided this interval into 1600
points.

\begin{figure}
\hspace*{2 cm}
\epsfig{file=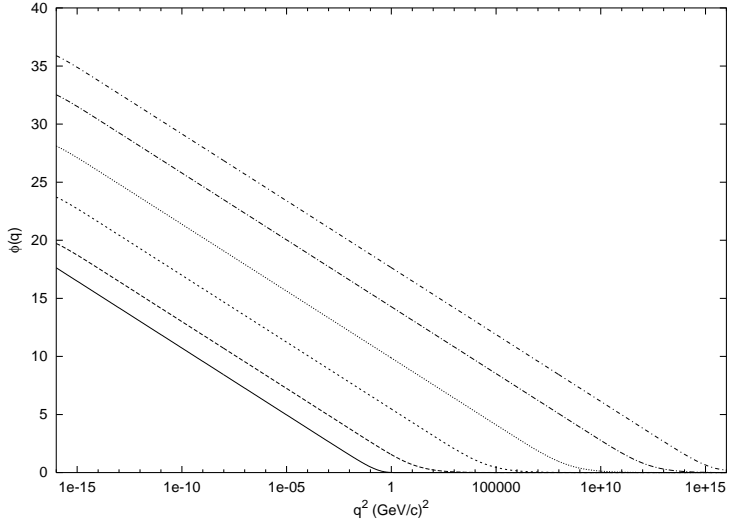,width=10 cm}
\caption{Solution $\phi(y,q)$ of Eq. (\ref{bkmom}) as a function of $q^2$
for different of $Y=\bar{\alpha}y$.
Curves from bottom to top correspond to $Y=0,2,4,6,8$ and 10.}
\label{fig5}
\end{figure}

The behaviour of $\phi(k)$
with $k^2$ at different values of the scaled rapidity
$Y=\bar{\alpha}y=2,4,6,8$ and 10
is shown in Fig. \ref{fig5}.
(Note that the maximal value of the scaled rapidity $Y=10$ corresponds
to the natural rapidity of order 50).

For the BFKL evolution in the external field $\phi$, in the first run (A),
we have taken the same form of the initial condition at $y=0$ but with
a variable slope
\beq
       p_0(k)=-(1/2){\rm Ei}(-k^2/a).
\label{ini1}
\eeq
We have performed calculations for $a=0.2,\  0.6,\ 1.0,\ 1.4$ and 1.8.
In the second run (B) the initial condition was taken with extra powers
of $k^2$.
\beq
       p_0(k)=-(1/2)k^{2n}{\rm Ei}(-k^2/0.3657)
\label{ini2}
\eeq
with $n=0,1,2,3$ and 4.

In all cases the behaviour of the solution $p(y,k)$ was found to be
universal. At large enough $y$ the solution becomes independent of $k^2$
up to a certain maximal $k^2_{max}(y)$, starting from which it goes to zero.
Roughly
\beq
p(y,k)\sim A(y)\theta(k^2_{max}-k^2).
\label{pyk}
\eeq
As $y$ grows $A(y)$ goes to zero  and $k^2_{max}(y)$ goes to infinity.
So on the whole the solution vanishes as $y\to\infty$, its $x$ dependence
tending to $\delta^2(x)$.

\begin{figure}
\hspace*{2 cm}
\epsfig{file=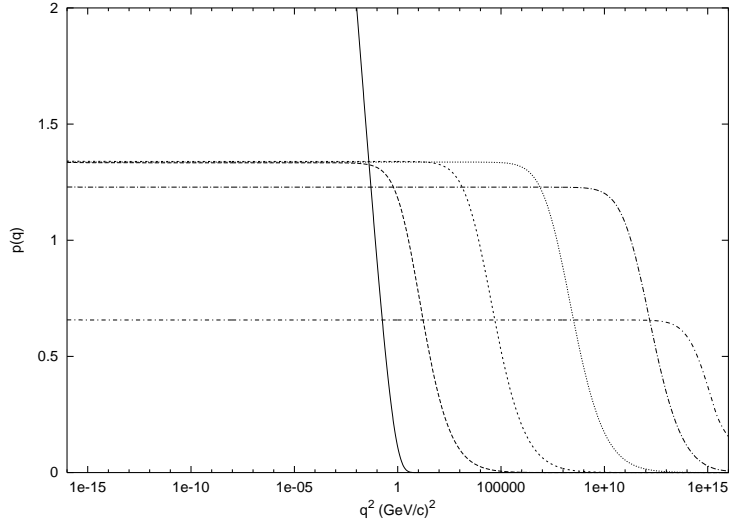,width=10 cm}
\caption{Solutions $p(y,q)$ of Eq. (\ref{eq1mom}) as a function of $q^2$
for different $Y=\bar{\alpha}y$ for run A with $a=1$. Curves which
start to fall at higher $q^2$
correspond to higher $Y=0,2,4,6,8$ and 10}
\label{fig6}
\end{figure}

\begin{figure}
\hspace*{2 cm}
\epsfig{file=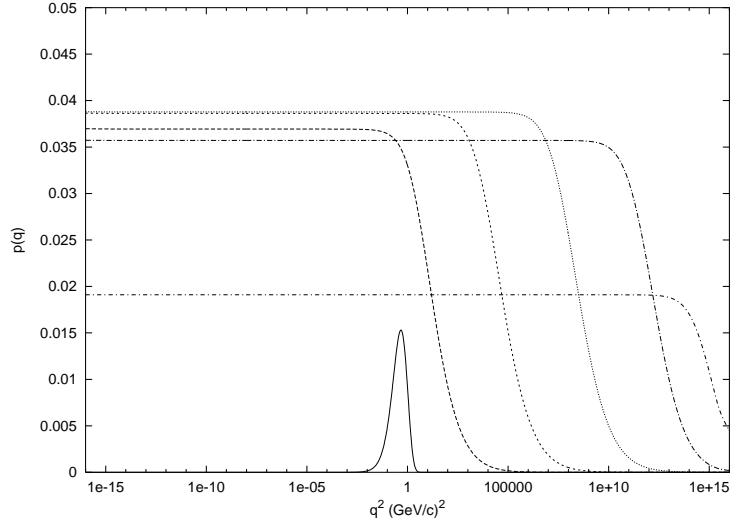,width=10 cm}
\caption{Solutions $p(y,q)$ of Eq. (\ref{eq1mom}) as a function of $q^2$
for different $Y=\bar{\alpha}y$ for run B with $n=2$. Curves which start to fall
at higher $q^2$ correspond to higher $Y=0,2,4,6,8$ and 10}
\label{fig7}
\end{figure}

We illustrate this behaviour in Figs. \ref{fig6} and \ref{fig7},
in which we show the solution $p(y,k)$ for run A with $a=1.0$ and run B with $n=2$ as a function
of $k^2$.
One observes that although the values of $p(y,k)$ for the two cases are
different, their behavior with $y$ is the same: they vanish as $y\to\infty$.

\begin{figure}
\hspace*{2 cm}
\epsfig{file=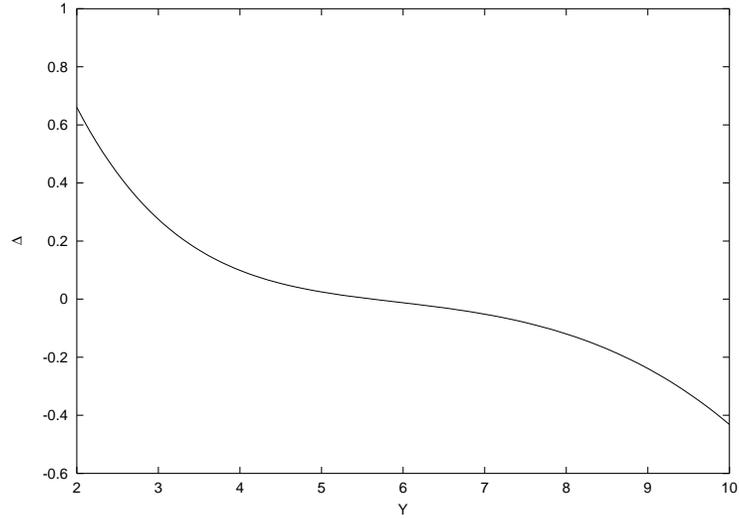,width=10 cm}
\caption{Slope $\Delta$ of the $Y$-dependence of the solutions $p(y,q)$
of Eq. (\ref{eq1mom}). The slope is the same for all initial conditions
(\ref{ini1}) and (\ref{ini2}).}
\label{fig8}
\end{figure}

This universality is especially obvious if one calculates the slope
$\Delta(y,k)$ of the $y$-dependence of
$p(y,k)$ at fixed $k$ presenting
\beq
p(y,k)\propto e^{Y\Delta(y,k)}.
\eeq
It turns out that at $Y>1$ the slope $\Delta(y,k)$ is independent of $k$ and
identical for all considered cases (run A with all studied $a$ and run B
with all studied $n$). Its smooth behaviour with $Y$ is shown in Fig.
\ref{fig8}. One observes that starting from $Y=5$ the slope becomes negative
indicating that the solution goes to zero at $Y>>1$.

\begin{figure}
\hspace*{2 cm}
\epsfig{file=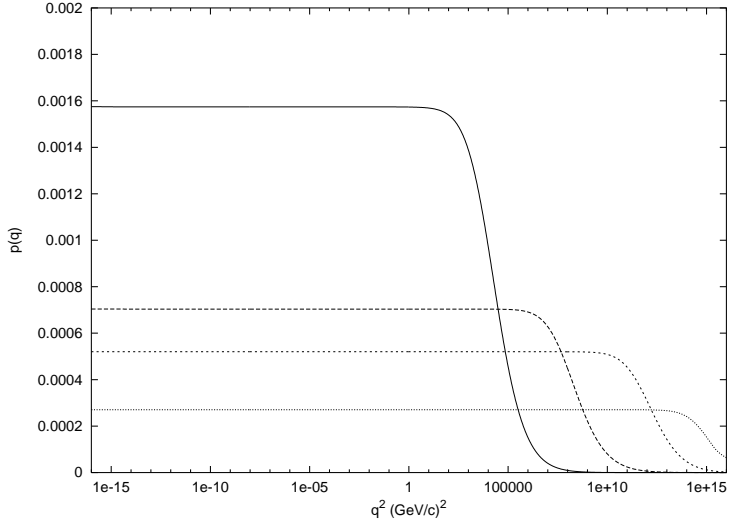,width=10 cm}
\caption{Solutions $p(y,q)$ of Eq. (\ref{eq1mom}) as a function of $q^2$
for different $Y=\bar{\alpha}y$ for the initial value taken at
$Y=3$ according to run A with $a=1$. Curves which
start to fall at higher $q^2$
correspond to higher $Y=4,6,8$ and 10}
\label{fig9}
\end{figure}

\begin{figure}
\hspace*{2 cm}
\epsfig{file=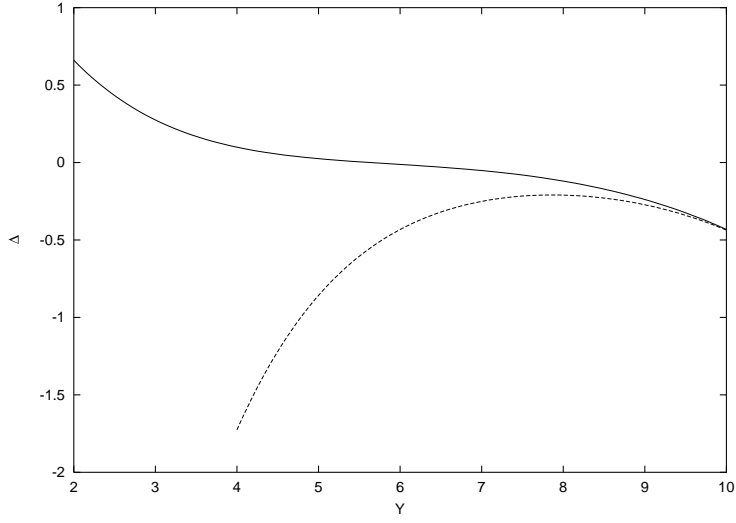,width=10 cm}
\caption{Slope $\Delta$ of the $Y$-dependence of the solutions $p(y,q)$
of Eq. (\ref{eq1mom}) with the initial condition put at $Y=0$ (upper curve)
and at $Y=3$ (lower curve) according to run A with $a=1$.}
\label{fig10}
\end{figure}

One has to take into account that in the external field $\phi(y,k)$
depending on rapidity the pomeron propagator ceases to depend only on the
rapidity difference. Rather the initial and final rapidities become two
independent variables. To see what influence it has on the behaviour of
the propagator at large rapidities we varied the initial rapidity $y=y'$
for the evolution of $p(y,k)$,  leaving unchanged the initial
rapidity $y=0$ for the evolution of $\phi(y,k)$,
which is the rapidity of the nucleus. One finds that although at initial
stages of evolution the behavior of $p(y,k)$ strongly depends on the
value of $y'$, at higher rapidities this behaviour is essentially the same
for any $y'$, namely the convoluted propagator goes down with rapidity
with the slope independent of $y_0$. This is illustrated in Figs. \ref{fig9}
and \ref{fig10} which show results for $y'=3/\bar{\alpha}\ (Y'=3)$
In Fig. \ref{fig9} we show the solution $p(y,k)$ for run A
with $a=1$.
One observes, that although absolute values of $p(y,k)$ are quite different
from the case $y'=0$ shown in Fig \ref{fig6}, the behaviour with the
growth of rapidity is the same. It is especially clear from the values
for the slope $\Delta$ shown in Fig. \ref{fig10} together with those for
the case $y=0$ (Fig. \ref{fig8}). Again at the initial stage of the evolution
the behavior with $Y'=3$ is quite different from that with $Y_0=0$.
However at higher rapidities the values for the slope become the
same.

\begin{figure}
\hspace*{2 cm}
\epsfig{file=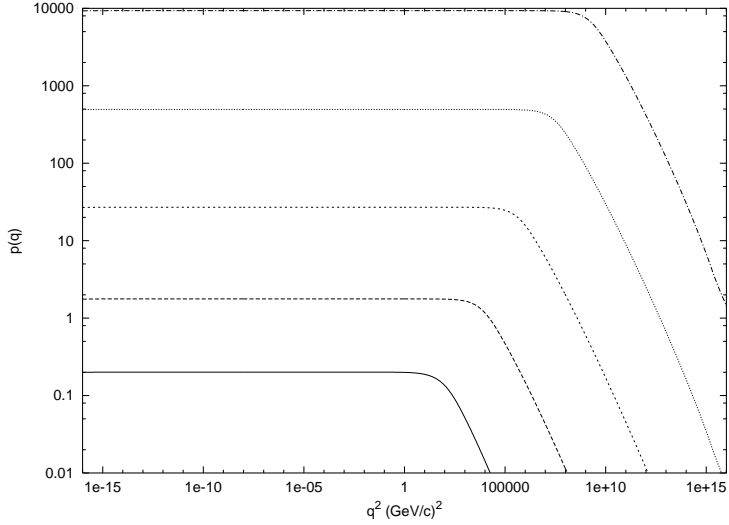,width=10 cm}
\caption{Solutions of Eq. (\ref{evoleq}) as a function of $k^2$ for different
$y$. Curves from bottom to top correspond to
$Y=\bar{\alpha}y=2,4,6,8$ and 10.}
\label{fig11}
\end{figure}

It is remarkable that this behaviour takes place only with $\phi$
given by the exact solution of the BK equation.
Taking an approximate form
\beq
\Phi(y,x)\simeq 1-e^{-Q^2(y)x^2},
\label{apprphi}
\eeq
where the "saturation momentum" $Q^2(y)\sim e^{2.05\bar{\alpha}y}$
we obtain an equation for $p(y,k)$ in the momentum space
\beq
\frac{\partial p(y,k)}{\partial y}=-\bar{\alpha}\Big[H_{BFKL}-
{\rm Ei}\Big(-\frac{k^2}{4Q^2(y)}\Big)\Big]p(y,k).
\label{evoleq}
\eeq
Taking for simplicity the initial condition $p_0(k)=\phi_0(k)$ at $y=0$
we get the solution shown in Fig. \ref{fig11}.
One observes that at large $y$ the solution acquires the same form
(\ref{pyk}) where however $A(y)$ grows with $y$:
\beq
A(y)\sim e^{1.4 Y}.
\eeq
This implies that with the approximate form (\ref{apprphi}) of $\Phi$
the final solution $P(y,x)$ in the coordinate space
behaves in a singular manner at $y\to\infty$. Effectively
\beq
P(y,x)_{y\to\infty}\to e^{1.4 y}x^2\delta^2(x)
\eeq
and it is impossible to say that it vanishes in this limit.

\subsection{Numerical studies. Coordinate space}
It is instructive to compare the behaviour of the propagator
in the momentum space with that in the coordinate space.
Rather than perform the Fourier transformation of the found
propagator $p(y,k)$ and determine $P(y,x)$ according to
(\ref{transmom}) we directly solved the forward equations
in the coordinate space (\ref{eqf}) and (\ref{ebkf}) choosing
the initial conditions in accordance with
(\ref{inphi}) and (\ref{ini1}).
\beq
\Phi(y=0,x)=1-e^{-\frac{1}{4}0.3657 x^2} \ \
P(y=0,x)=1-e^{-\frac{1}{4}a x^2}.
\label{inix}
\eeq
Again we used the Runge-Kutta method with $10^{-8}<r<10^8$. The number
of divisions was taken 400 in $\ln r^2$ and 200  in the azimuthal angle.

\begin{figure}
\hspace*{2 cm}
\epsfig{file=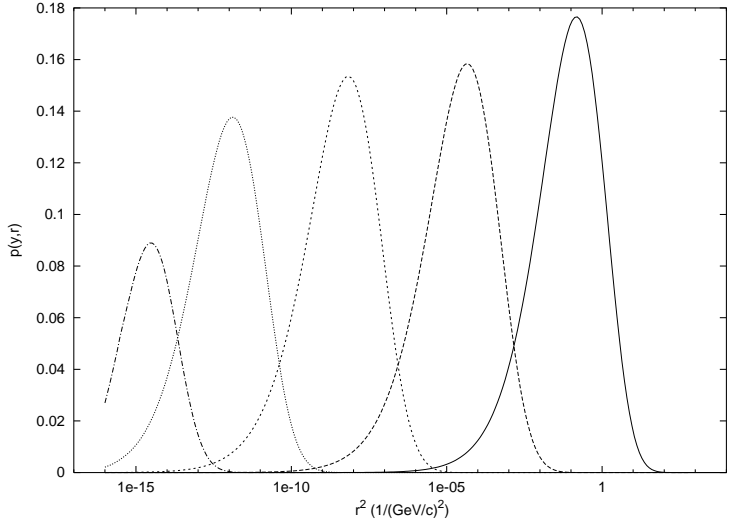,width=10 cm}
\caption{Solutions of Eq. (\ref{eqf}) in the nuclear field as a
function of $x^2$ for different
$y$. Curves with a maximum from right to left  correspond to
$Y=\bar{\alpha}y=0,2,4,6,8$ and 10.}
\label{fig12}
\end{figure}

 Numerical calculations
become much more cumbersome in the coordinate space due to the
fact that the non-linear term and the term with the nuclear
field depend on two different points. Having in mind that these
calculations serve only for the illustrative purpose, we have restricted
ourselves to only one value $a=0.3657$ in (\ref{inix}), that is
have taken the initial values for $P(y,x)$ at $y'=0$
the same as for $\Phi(y,x)$.
Our results are shown in Fig. \ref{fig12}.
The form of the solution in the coordinate space very much resembles that
of the gluon density in the momentum space: there is a sharp maximum at
a certain value of $x^2$ which with the growth of $y$ moves  in the direction
of smaller $x^2$ (to the ultraviolet). However in contrast to the
gluon density the height of the maximum  is diminishing with $y$, so that
in the limit $y\to\infty$ the propagator vanishes.

\begin{figure}
\hspace*{2 cm}
\epsfig{file=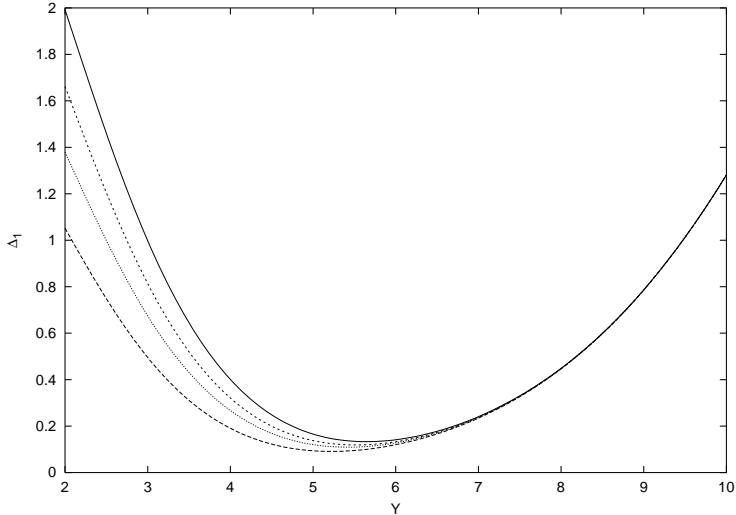,width=10 cm}
\caption{Exponent $\Delta_1$ in (\ref{conv1}) as a function of rapidity
for the convolution of solutions to Eq. (\ref{eqf}) with the
initial conditions in the form (\ref{inix}). The curves from top to bottom
correspond to 1) all three solutions with $a=0.3657$, 2) two with
$a=0.3657$ and one with $a=1.4$, 3) two with  $a=1.4$ and one with
$a=0.3657$ and the lowermost to all three with $a=1.4$.}
\label{fig13}
\end{figure}

In the construction of the amplitudes  pomeron propagators are convoluted
with the triple pomeron vertex. The behaviour of this convolution with
the growth of rapidities can be studied from the integral
\beq
I_1(y)=\int \frac{d^2x_1d^2x_2}{x_1^2x_2^2}P_1(y,x_1)P_2(y,x_2)P_3(y,x_1+x_2)
\eeq
where $P_{1,2,3}$ may correspond to the same or different initial conditions.
Presenting $I_1$ as
\beq
I_1(y)=a_1e^{-\Delta_1(y) Y},\ \ y=\bar{\alpha}Y
\label{conv1}
\eeq
we find values of the exponent $\Delta_1$ from our numerical calculations
shown in Fig. \ref{fig13} for the initial conditions (\ref{inix}) with
$a=0.3657$ and $a=1.4$. As one observes, although at lower  rapidities $Y$
the values of $\Delta_1$ depend on the choice of the initial conditions,
starting from $Y=6$ this dependence disappears and the behaviour of
$\Delta_1$ becomes universal. In any case it is clear that the convolution
vanishes at $Y\to\infty$. To compare, it grows as $\exp (3\Delta_{BFKL}Y)$
in the vacuum, with $\Delta_{BFKL}=4\ln 2$.

\section{General case}
Calculations of the loop contribution require the propagator in the
non-forward direction, depending not only on the relative distance between
the gluons but also on the center-of mass coordinate.The convoluted propagator
$P(y,r_1,r_2)$ in this case satisfies Eq. (\ref{eqq}) with the initial
condition (\ref{ini}). As to the nuclear field $\Phi(y,r_1,r_2)$,
we may continue to consider it taken in the forward direction,
neglecting the
small nuclear momenta transferred to the gluons. So $\Phi$ continues
to depend only on $r_{12}$ and satisfies the same equation (\ref{ebkf})
and initial condition (\ref{inix}).

Propagator $P(y,r_1,r_2)$, apart from $y$ depends on three variables
for which we may take $r_1^2$, $r_2^2$ and angle $\phi$ between $r_1$
and $r_2$. We introduce a grid in variables $\ln r_1^2$, $\ln r_2^2$
and $\phi$, dividing the interval in the first two variables into $N_1$
points and in $\phi$ into $N_2$ points.  The evolution equation
(\ref{eqq}) then  requires calculation of the
integrand at $N_1^3\cdot N_2^2$ points at each value of intermediate
rapidities. This severely restricts the numbers
of $N_1$ and $N_2$ admissible for given calculation facilities.
As a result, in the past there were quite few calculations
of the solution to the non-forward BK equation ~\cite{GBS,berger}
with modest values of $N_1$ and $N_2$. In our case we use the Runge-Kutta
method of the solution with the maximal value $N_1=N_2=80$ compatible with
the reasonable calculation time. The limits in $r$ were taken as before
$10^{-8}<r<10^8$.
The initial condition (\ref{ini}) for the convoluted propagator
was taken at $y'=0$ in the form borrowed from ~\cite{GBS}
\beq
P(y=0,r_1,r_2)=1-e^{-c_1r_{12}^2e^{-b^2/c_2}},
\label{inii}
\eeq
where $b=(r_1+r_2)/2$. We have studied 9 cases with $c_1=0.1,\,1$ and 10
and $c_2=0.2,\, 2$ and 20.
In all cases the results are quite similar.

\begin{figure}
\hspace*{2 cm}
\epsfig{file=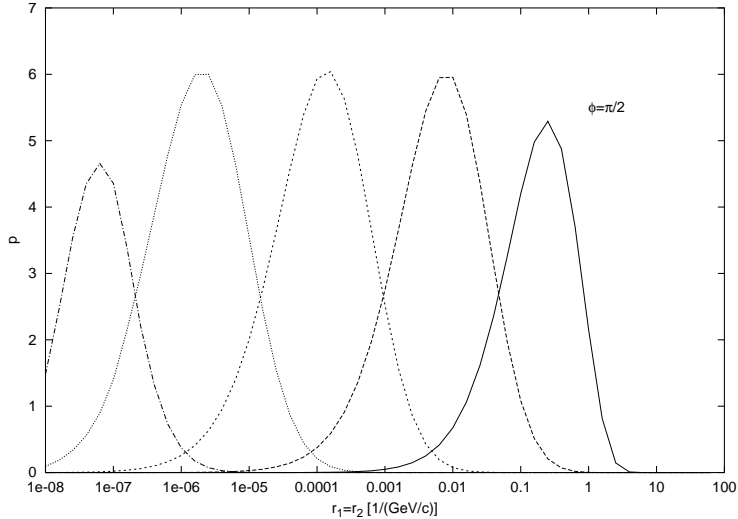,width=10 cm}
\caption{Propagator $P(y,r_1=r_2,\phi=\pi/2)$ as a function of $Y$
for the initial conditions (\ref{inii}) with $c_1=10$ and $c_2=2$.
Curves with maxima from right to left correspond to $Y=2,4,6,8,10$
respectively.}
\label{fig14}
\end{figure}

Calculations produce an array of data $P(y,r_1^2,r_2^2,\phi)$
on the three-dimensional grid. The results can be presented in various ways.
In Fig. \ref{fig14} we show values of
$P(y,r_1^2,r_2^2,\phi)$ with $c_1=10$ and $c_2=2$
at $r_1=r_2$ and $\phi=\pi/2$ for different $y$.

As we see the general trend of the curves is the same as for the forward
case, Fig \ref{fig12}, although the decrease of the height of the maximum
starts from larger rapidities. The behaviour of $P(y, r_1=r_2,\phi)$
at different values of $\phi$ and for different $c_1$ and $c_2$ in the
initial conditions (\ref{ini}) is the same, although the value
of the rapidity $Y$ at which the decrease starts is slightly different
for different initial conditions.

\begin{figure}
\hspace*{2 cm}
\epsfig{file=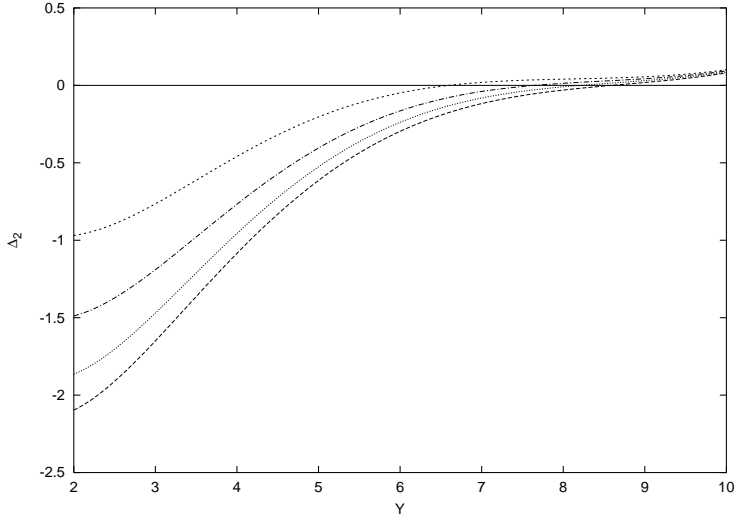,width=10 cm}
\caption{Exponent $\Delta_2$ in (\ref{conv2}) as a function of rapidity
for the convolution of solutions to Eq. (\ref{eqq}) with the
initial conditions (IC) A and B in (\ref{inii})(see explanation in the text.
The curves from top to bottom
correspond to 1) all three solutions with IC=A  2) two with
IC=A  and one with B 3) two with  IC= B
and one with A and the lowermost to all three solutions with
IC=B}
\label{fig15}
\end{figure}

To have an overall picture of the behaviour of the non-forward propagator
with rapidity, as for the forward case, we studied the convolution
of three propagators with the triple pomeron vertex (dropping operator $K$, which hopefully does not change the behaviour
at large $y$, see ~\cite{bra4})

\beq
I_2(y)=\int\frac{d^2r_1d^2r_2d^2r_3}{r_{12}^2r_{23}^2r_{31}^2}
P_1(y, r_1,r_2)P_2(y,r_2,r_3)P_3(y,r_3,r_1),
\label{conv2}
\eeq
where $P_1,\ P_2$ and $P_3$ may come from the same or different
initial conditions. We again present
\beq
I_2(y)=a_2e^{-\Delta_2(y) Y},\ \ y=\bar{\alpha}Y
\eeq
and in Fig. \ref{fig15} show values of $\Delta_2(y)$ for four cases,
with the three propagators taken 1) all for the initial condition (IC)
with $c_1=10,\ c_2=2$ (IC=A) 2)all for the IC with $c_1=1,\ c_2=20$ (IC=B)
3)two propagators with
IC=A and one with IC=B and 4) two propagators with IC=B and one with IC=A.
We see that at low values of $Y$ the exponents $\Delta_2$ are different for
these cases and predominantly negative, which means that the propagators
in fact grow at these rapidities. However at larger rapidities
different $\Delta_2$ converge to a common positive value indicating that the
convolution decreases with rapidity and that this decrease is universal.


\section{Conclusions}
We have studied numerically the BFKL pomeron propagator in
the external field created by the solution of the BK
equation in the nuclear matter. We have found that for more or less
arbitrary set of initial conditions the convoluted propagator
vanishes at large rapidities, its coordinate dependence in the forward
case tending to the $\delta$-function.  This gives reasons
to believe that the  propagator itself vanishes at large rapidities
in the nuclear background. This result follows only with the field being the
exact solution of the BK equation.

Our results may be a starting point for the perturbative calculation of loop
contributions in the nuclear background.
However the technical difficulties seem to be quite formidable, since
calculations are only possible numerically and even  simplest loops
involve  quite a number of spatial points over which one has to integrate.
One needs to invent a reliable approximate representation for the
non-forward propagator in the nuclear field which would allow one
to perform all these integrations effectively.
A somewhat simpler application is to the study of double inclusive cross-section
for gluon jet production in nucleus-nucleus collisions, which only requires
the forward propagator. Both these complicated problems are left for
future investigations.

\section{Acknowledgments}
The authors are indebted to J.Bartels, E.Levin, L.Lipatov and A.Mueller
for discussion. This work has been supported by grants RFFI 09-012-01327-a and RFFI-CERN
08-02-91004. A.T. also acknowledges the FTP "Scientific and scientific-pedagogical personnel of innovative Russia in 2009-2013" (Contract No. P20).

\end{document}